\documentclass[12pt]{article}
\usepackage[dvips]{graphicx}
\usepackage{amssymb}
\usepackage[hang]{caption2}
\usepackage{psfrag}

\def\dspace{\baselineskip = 0.25in}

\begin{document}

\dspace
\begin{titlepage}
\begin{flushright}
BA-03-05\\
\end{flushright}
\vskip 2cm
\begin{center}
{\Large\bf
Testing Supersymmetric Grand Unified \\ Models of Inflation
}
\vskip 1cm
{\normalsize\bf
V. N. Senoguz\footnote{nefer@udel.edu} and
Q. Shafi\footnote{shafi@bxclu.bartol.udel.edu}
}
\vskip 0.5cm
{\it Bartol Research Institute, University of Delaware, \\Newark,
DE~~19716,~~USA\\[0.1truecm]}

\end{center}
\vskip .5cm


\begin{abstract}
We reconsider a class of well motivated supersymmetric models in which inf\mbox{}lation is associated
 with the breaking of a gauge symmetry $G$ to $H$, with the symmetry breaking
 scale $M\sim10^{16}$ GeV. Starting with a renormalizable superpotential, we include
 both radiative and supergravity corrections to derive the inf\mbox{}lationary
 potential. The scalar spectral index $n_{s}$ can exceed unity in some cases,
 and it cannot be smaller than 0.98 if the number of e-foldings 
corresponding to the present horizon scale is around 60.
Two distinct variations of this scenario are discussed in which non-renormalizable terms 
allowed by the symmetries are included in the superpotential, and one finds $n_s\geqslant0.97$.
 The models discussed feature a tensor to scalar ratio $r\lesssim10^{-4}$, while $\textrm {d}n_{s}/\textrm {d}\ln k\lesssim10^{-3}$. If $G$
 corresponds to $SO(10)$ or one of its rank five subgroups, the observed baryon asymmetry
 is naturally explained via leptogenesis. 

\end{abstract}
\end{titlepage}
\newpage

\section{Introduction}
 Supersymmetric grand unified theories in four and higher dimensions continue
 to play a prominent role in high energy physics, and it is therefore tempting
 to speculate that they may also play a key role in realizing an inf\mbox{}lationary
 epoch in the very early universe. Indeed, in a class of realistic supersymmetric models, 
inf\mbox{}lation is associated with the breaking either of a grand
 unified symmetry or one of its subgroups. 

In the simplest models, inf\mbox{}lation
 is driven by quantum corrections generated by supersymmetry breaking in the
 early universe, and the temperature fluctuations $\delta T/T$
 are proportional to $(M/M_{P})^2$, where $M$ denotes the symmetry breaking
 scale of $G$, and $M_{P}=1.2\times10^{19}$ GeV denotes the Planck mass \cite{dvaliet.,lazarides}. It turns out
 that for $M\sim10^{16}$ GeV, one predicts an essentially scale invariant spectrum which is consistent with a variety of 
CMB measurements 
 including the recent WMAP results \cite{wmap}. With inf\mbox{}lation `driven' solely by radiative corrections the scalar spectral
 index $n_{s}$ is very close to $0.98$, if the number of e-foldings $N_Q$ 
after the present horizon scale crossed outside the inf\mbox{}lationary horizon is close \mbox{to 60}.

As an example, if $G=SO(10)$, one
 could associate inf\mbox{}lation with the breaking of $SO(10)$ to $SU(5)$. A realistic
 model along these lines is most easily realized in a five dimensional
 setting \cite{kyaeet.}, in which compactification on an orbifold can be exploited
 to break $SO(10)$ down to the MSSM. Interesting examples for $G$ in four
 dimensions include the gauge symmetry $SU(4)_{c}\times SU(2)_{L}\times SU(2)_{R}$ \cite{patisalam,king,jeannerot}
 as well as $SU(2)_{L}\times SU(2)_{R}\times U(1)_{B-L}$ \cite{lss,dls}. 
 If the unified gauge group $G$ is identified with $SO(10)$ or one of its subgroups
 listed above, the inf\mbox{}laton naturally decays into massive right-handed
 neutrinos whose out of equilibrium decay lead to the observed baryon
 asymmetry via leptogenesis.

Motivated by the prospects that much more precise information about the
 scalar index $n_{s}$ and other related quantities will become available in the
 not too distant future, especially from a successful launch of the Planck satellite, we reconsider
 a realistic set of inf\mbox{}lationary models in which not only radiative
 but also the canonical supergravity (SUGRA) contribution to the inf\mbox{}lationary potential is taken into
 account. As noted in \cite{panagio,linde,kawasaki}, the presence of SUGRA corrections can give
 rise to $n_{s}$ values that exceed unity. We find that $n_{s}$ does indeed exceed unity 
for some parameter choices, and its value cannot
 fall bellow $0.97$ for e-foldings close to $60$. 

There are no tiny dimensionless parameters in
 the class of models that we consider. Indeed, in the tree level superpotential there 
appears a dimensionless coupling $\kappa$ whose value, it turns
 out, is restricted to be $\lesssim0.1$.
Otherwise the scalar spectral index exceeds unity due to SUGRA corrections by an amount that is not favored
by the data on smaller scales. With $\textrm{d}n_{s}/\textrm{d}\ln k\lesssim10^{-3}$, this requires that the vacuum energy density
that drives inf\mbox{}lation is somewhat below $M^4_{GUT}$, and the tensor to scalar ratio $r$
turns out to be $\lesssim10^{-4}$. 

The plan of the paper is as follows. In section 2 we discuss the simplest supersymmetric GUT inflation model 
realized with a renormalizable superpotential whose form is fixed by a $U(1)$ R-symmetry.
The termination of inf\mbox{}lation in this case is abrupt, leading to a monopole problem for models 
such as $SU(4)_{c}\times SU(2)_{L}\times SU(2)_{R}$. (For a resolution in $SU(5)$ see \cite{covi}.)
In sections 3 and 4,
we consider two extensions of this scenario in which non-renormalizable terms allowed by the symmetries are included in the superpotential,
and the monopole problem is circumvented. In one extension, 
called `shifted' GUT inf\mbox{}lation \cite{jeannerot}, the gauge symmetry is already
broken along the inf\mbox{}lationary trajectory. In the other extension, called `smooth' GUT inf\mbox{}lation \cite{smooth}, 
the inf\mbox{}lationary path possesses a classical inclination and the termination of inf\mbox{}lation is smooth.
Finally, in section 5 we note that consideration of leptogenesis in an $SO(10)$ model results in a somewhat more stringent
upper bound on $\kappa\,(\lesssim10^{-2})$. 

\section{Supersymmetric GUT Inflation} 
An elegant inf\mbox{}lationary scenario is most readily
realized starting with the renormalizable superpotential \cite{dvaliet.,copeland}

\begin{equation} \label{super}
W_1=\kappa S(\phi\overline{\phi}-M^{2})
\end{equation}

\noindent where $\phi(\overline{\phi})$ denote a conjugate pair of superfields transforming as nontrivial representations of
some gauge group $G$, $S$ is a gauge singlet superfield, and $\kappa$ $(>0)$ is a dimensionless coupling. 
A suitable $U(1)$ R-symmetry, under which $W_1$ and $S$ transform the same way, ensures the uniqueness of this superpotential
at the renormalizable level \cite{dvaliet.}.  
In the absence of supersymmetry breaking, the potential energy minimum corresponds to non-zero (and equal in magnitude) vevs $(=M)$
for the scalar components in $\phi$ and $\overline{\phi}$, while the vev of $S$ is zero. (We use the same notation for superfields
and their scalar components.)
Thus, $G$ is broken to some subgroup $H$. In the presence of $N=1$
supergravity, $S$ acquires a vev comparable to the gravitino mass $m_{3/2}$ ($\sim$ TeV).

In order to realize inf\mbox{}lation, the scalar fields $\phi$, $\overline{\phi}$, $S$ must be displayed from their present minima.
Thus for $|S|>M$, the $\phi$, $\overline{\phi}$ vevs both vanish so that the gauge symmetry is restored, and the tree level
potential energy density $\kappa^{2}M^{4}$ dominates the universe. With supersymmetry thus broken, there are radiative
corrections from the $\phi-\overline{\phi}$ supermultiplets that provide logarithmic corrections to the potential
which drives inf\mbox{}lation. In one loop approximation the inf\mbox{}lationary effective potential is given by \cite{dvaliet.}

\begin{equation} \label{loop}
V_{\mathrm{LOOP}}=\kappa^{2}M^{4}\left[1 +\frac{\kappa^{2}\mathcal{N}}{32\pi^{2}} \left(
2\ln\frac{\kappa^{2}|S|^{2}}{\Lambda^{2}}+(z+1)^{2}\ln(1+z^{-1})+ (z-1)^{2}\ln(1-z^{-1})\right) \right]\,,
\end{equation}

\noindent where $z=x^{2}=|S|^{2}/M^{2}$, $\mathcal{N}$ is the dimensionality of the representations to which
$\phi$, $\overline{\phi}$ belong, and $\Lambda$ is a renormalization mass scale. From Eq.\,(\ref{loop}) the
quadrupole anisotropy is found to be \cite{dvaliet.,lazarides,lss}:

\begin{equation} \label{quad}
\left(\frac{\delta T}{T}\right)_{Q}\approx\frac{8\pi}{\sqrt{\mathcal{N}}}\left(\frac{N_{Q}}{45}\right)^{\frac{1}{2}}
\left(\frac{M}{M_{P}}\right)^{2}x_{Q}^{-1}y_{Q}^{-1}f (x^{2}_{Q})^{-1}\,,
\end{equation}

\noindent with

\begin{equation}
f(z)=\left(z+1\right)\ln\left(1+z^{-1}\right)+\left(z-1\right)\ln\left(1-z^{-1}\right)\,,
\end{equation}

\begin{equation} \label{yq}
y_{Q}^{2}=\int_{1}^{x^{2}_{Q}}\frac{\textrm{d} z}{z f(z)}\quad ,y_Q\ge 0\,.
\end{equation}

\noindent Here, the subscript $Q$ denotes the epoch when the present horizon scale crossed outside the inf\mbox{}lationary horizon
and $N_{Q}$ is the number of e-foldings it underwent during inf\mbox{}lation. From Eq.\,(\ref{loop}), one also obtains

\begin{equation} \label{m_kap}
\kappa\approx\frac{8\pi^{3/2}}{\sqrt{\mathcal{N}N_{Q}}}\,y_{Q}\,\frac{M}{M_{P}}\,.
\end{equation}

\noindent For relevant values of the parameters ($\kappa\ll1$), the slow roll conditions are violated only `infinitesimally' close
to the critical point at $x=1$ ($|S|=M$) \cite{lazarides}. So inf\mbox{}lation continues practically until this point is reached,
where the `waterfall' occurs.

Several comments are in order:

\begin{itemize}
\item For $x_{Q}\gg1$ (but $|S_{Q}|\ll M_{P}$), $y_{Q}\to x_{Q}$ and $x_{Q}\,y_{Q}\,f(x^{2}_{Q})\to1^{-}$.

\item Comparason of Eq.\,(\ref{quad}) with the COBE result $(\delta T/T)_{Q}\simeq6.3\times10^{-6}$ \cite{cobe} shows that
the gauge symmetry breaking scale $M$ is naturally of order $10^{16}$ GeV.

\item Suppose we take $G=SO(10)$, with $\phi(\overline{\phi})$ belonging to the $\textbf{16}(\overline{\textbf{16}})$
representation, so that $G$ is spontaneously broken to $SU(5)$ at scale $M$. Taking $\mathcal{N}=16$, and
 $x_{Q}\,y_{Q}\,f(x^{2}_{Q})\to1^{-}$, $M$ is determined to be $10^{16}$ GeV, which essentially coincides with the
SUSY GUT scale. The dependence of $M$ on $\kappa$ is displayed in Fig.\,1. Note that a five dimensional supersymmetric
$SO(10)$ model in which inf\mbox{}lation is associated with this symmetry breaking was presented in \cite{kyae}. 

\item Another realistic example is given by
$G=SU(3)_c\times SU(2)_{L}\times SU(2)_{R}\times U(1)_{B-L}$, corresponding to $\mathcal{N}=2$, and the
scale $M$ is then associated with the breaking of $SU(2)_{R}\times U(1)_{B-L}\to U(1)_{Y}$ \cite{lss,dls}.

\item The scalar spectral index $n_s$ is given by \cite{ll}

\begin{equation}
n_s\cong1-6\epsilon+2\eta,\qquad\epsilon\equiv \frac{m^2_P}{2}\left(\frac{V'}{V}\right)^2,\qquad\eta\equiv\frac{m^2_P V''}{V},
\end{equation}

\noindent where $m_P$ is the reduced Planck mass $M_P/\sqrt{8\pi}$; hereafter we take $m_P=1$. The primes denote derivatives with 
respect to the normalized real scalar field $\sigma\equiv\sqrt{2}|S|$. For $x_{Q}\gg1$ (but $\sigma_{Q}\ll1$), $n_s$
approaches \cite{dvaliet.}
 
\begin{equation}
n_{s}\simeq1+2\eta\simeq 1-\frac{1}{N_{Q}}\simeq0.98
\end{equation}

where $N_{Q}\approx60$ denotes the number of e-foldings.\footnote{
$N_{Q}\simeq56.5+(1/3)\ln(T_r/10^9\,\textrm{GeV})+(2/3)\ln(\mu/10^{15}\,\textrm{GeV})$ \cite{lazarides}, 
where $T_r$ is the reheating temperature
and $\mu$ is the false vacuum energy density.} The dependence of $n_{s}$ on $\kappa$ is displayed in Fig.\,2 (the behavior of
$n_s$ for large $\kappa$ is inf\mbox{}luenced by the SUGRA correction, as discussed below).

\item The minimum number of e-foldings ($\approx60$) required to solve the horizon and flatness problems can be achieved
even for $x_{Q}$ very close to unity, provided that $\kappa$ is taken to be sufficiently small. This follows from Eq.\,(\ref{m_kap}).
An important constraint on $\kappa$ can arise from considerations of the reheat temperature $T_{r}$ after inf\mbox{}lation,
taking into account the gravitino problem. The latter requires that $T_{r}\lesssim10^{10}$ GeV \cite{gravitino}, unless
some mechanism is available to subsequently dilute the gravitinos. 

The inf\mbox{}laton mass is $\sqrt{2}\kappa M$ (recall that both $S$, and $\phi$, $\overline{\phi}$ oscillate about their
minima after inf\mbox{}lation is over, and they have the same mass), and so to prevent inf\mbox{}laton decay via gauge interactions
which would cause $T_r$ to be too high ($\sim M\gg10^{10}$ GeV), the coupling $\kappa$ should not exceed unity.
A more stringent constraint on $\kappa\ (\leqslant0.1)$ appears when SUGRA corrections are included.

\item For $G=SO(10)$ or $SU(3)_c\times SU(2)_{L}\times SU(2)_{R}\times U(1)_{B-L}$, 
the inf\mbox{}laton produces right handed neutrinos \cite{ls,lss,lsv,Pati:2002pe} whose subsequent out of
equilibrium decay leads to the observed baryon asymmetry via leptogenesis \cite{lepto,ls}.

\item For sufficiently large values of $\kappa$, SUGRA corrections become important, and more often than not, these tend to derail
an otherwise succesful inf\mbox{}lationary scenario by giving rise to scalar mass$^2$ terms of order $H^2$, where $H$ denotes
the Hubble constant. Remarkably, it turns out that for a canonical SUGRA potential (with minimal K\"ahler potential
$|S|^2+|\phi|^2+|\overline{\phi}|^2$), the problematic mass$^2$ term cancels out for the superpotential $W_1$ in
Eq.\,(\ref{super}) \cite{copeland}. This may be considered an attractive feature of the inf\mbox{}lationary scenario. Note
that this property persists even when non-renormalizable terms that are permitted by the $U(1)_R$ symmetry are included
in the superpotential.

\end{itemize}

\noindent The SUGRA scalar potential is given by

\begin{equation} \label{sugra}
V=e^{K/m^2_P}\left[\left|\frac{\partial W}{\partial z_i}+\frac{z^*_i W}{m^2_P}\right|^2-3\frac{|W|^2}{m^2_P}\right]\,,
\end{equation}

\noindent where the sum extends over all fields $z_i$, and $K=\sum_i |z_i|^2$ is the minimal K\"ahler potential.
(In general, $K$ is expanded as $K=|S|^2+|\phi|^2+|\overline{\phi}|^2+\alpha|S|^4/m^2_P+\ldots$, and only the $|S|^4$ term
in $K$ generates a mass$^2$ for $S$, which would ruin inf\mbox{}lation for $\alpha\sim1$ \cite{Panagiotakopoulos:1997ej,Lazarides:1998zf}. 
From the requirement $\sigma<m_P$,
one obtains an upper bound on $\alpha\ (\lesssim10^{-2})$ \cite{Hamaguchi:2002vc}.)
From Eq.\,(\ref{sugra}), the SUGRA correction to the potential is \cite{copeland, panagio, linde, kawasaki}

\begin{equation}
V_{SUGRA}=\kappa^{2}M^{4}\left[\frac{1}{8}\sigma^{4}+\ldots\right]\,,
\end{equation}

\noindent where $\sigma=\sqrt{2}|S|$ is a normalized real scalar field, and we have set the reduced Planck mass $m_P=1$. 
The effective inf\mbox{}lationary potential $V_1$ can be written to a good approximation as the sum of the radiative and SUGRA corrections.
For $1\gg\sigma\gg \sqrt{2}M$,

\begin{equation} \label{v1}
V_1\approx\kappa^{2}M^{4}\left[1+\frac{\kappa^{2}\mathcal{N}}{32\pi^{2}}2\ln\frac{\kappa^{2}\sigma^{2}}{2\Lambda^{2}}+
\frac{1}{8}\sigma^{4}\right]\,,
\end{equation}

\noindent and comparing the derivatives of the radiative and SUGRA corrections one sees that the radiative term dominates for
$\sigma^2\lesssim\kappa\sqrt{\mathcal{N}}/2\pi$. From $3H\dot{\sigma}=-V'$, $\sigma^2_Q\simeq\kappa^2 \mathcal{N}N_Q/4\pi^2$ for the
one-loop effective potential, so that SUGRA effects are negligible only for $\kappa\ll2\pi/\sqrt{\mathcal{N}}N_{Q}\simeq
0.1/\sqrt{\mathcal{N}}$. (For $\mathcal{N}=1$, this essentially agrees with \cite{linde}).

\noindent From Eq.\,(\ref{v1}), the scalar spectral index is given by

\begin{equation}
n_s\simeq1+2\eta\simeq1+2\left(3\sigma^2-\frac{\kappa^2\mathcal{N}}{8\pi^2\sigma^2}\right)\,,
\end{equation}

\noindent and it exceeds unity for $\sigma^2\gtrsim\kappa\sqrt{\mathcal{N}}/2\sqrt{3}\pi$. For $x_Q\gg1$,

\begin{equation} \label{nq}
N_Q=\int_{\sigma_{end}}^{\sigma_Q}\frac{V}{V'}\textrm{d}\sigma
\approx\frac{\pi}{2\sigma^2_Q}\frac{\kappa}{\kappa_c}\tan\left(
\frac{\pi}{2}\frac{\kappa}{\kappa_c}\right)\,,
\end{equation}

\noindent where $\kappa_c=\pi^2/\sqrt{\mathcal{N}}N_Q\simeq0.16/\sqrt{\mathcal{N}}$.  
Using Eq.\,(\ref{nq}), one finds that the spectral index exceeds unity for $\kappa\simeq2\pi/\sqrt{3\mathcal{N}}N_Q\simeq0.06/\sqrt{\mathcal{N}}$.

The quadrupole anisotropy is found from Eq.\,(\ref{v1}) to be

\begin{equation}
\left(\frac{\delta T}{T}\right)_{Q}=\frac{1}{4\pi\sqrt{45}}\frac{V^{3/2}_1}{V'_1}\approx
\frac{1}{2\pi\sqrt{45}}\frac{\kappa\,M^2}{\sigma^3_Q}\,.
\end{equation}

\noindent In the absence of the SUGRA correction, the gauge symmetry breaking scale $M$ calculated from the observed quadrupole 
anisotropy approaches
the value $\mathcal{N}^{1/4}\cdot6\times10^{15}$ GeV for $x_Q\gg1$ (from Eq.\,(\ref{quad}), with $x_{Q}\,y_{Q}\,f(x^{2}_{Q})\to1^{-}$).
The presence of the SUGRA term leads to larger values of $\sigma_Q$ and hence larger values of $M$ for 
$\kappa\gtrsim0.06/\sqrt{\mathcal{N}}$. The dependence of $M$ on $\kappa$ including the full one-loop
potential (Eq.\,(\ref{loop})) and the leading SUGRA correction is presented in Fig.\,1.

To summarize, the
scalar spectral index in this class of models is close to unity for small $\kappa$, has a minimum 
at $\simeq0.98$ for
$\kappa\simeq0.02/\sqrt{\mathcal{N}}$, and exceeds unity for $\kappa\gtrsim0.06/\sqrt{\mathcal{N}}$ (Fig.\,2).
The experimental data seems not to favor $n_s$ values in excess of unity on smaller scales (say $k\sim0.05$ Mpc$^{-1}$),
which leads us to restrict ourselves to $\kappa\lesssim0.06/\sqrt{\mathcal{N}}$. 
Thus, even though the symmetry breaking scale $M$ is of order
$10^{16}$ GeV (Fig.\,1), the vacuum energy density during inf\mbox{}lation is smaller than $M^4_{GUT}$. Indeed,
the tensor to scalar ratio $r\lesssim10^{-4}$ (Fig.\,4). Finally, the quantity $\textrm{d}n_{s}/\textrm{d}\ln k$ is negligible
for small $\kappa$ and $\sim10^{-3}$ as the spectral index crosses unity \cite{kawasaki} (Fig.\,3). The WMAP team has reported a value for 
$\textrm{d}n_{s}/\textrm{d}\ln k=-0.042^{+0.021}_{-0.020}$ \cite{wmap}, 
but the statistical significance of this conclusion has been questioned by the authors of \cite{seljak}. Clearly, more data is
necessary to resolve this important issue. Modifications of the models discussed here has been proposed in \cite{kawasaki} to
generate a much more significant variation of $n_s$ with $k$.
 
\section{Shifted GUT Inflation}
The inf\mbox{}lationary scenario based on the superpotential $W_1$ in Eq.\,(\ref{super}) has the characteristic feature that the end of 
inf\mbox{}lation
essentially coincides with the gauge symmetry breaking $G$ (e.g. $SO(10)$) to $H$ $\left(SU(5)\right)$. Thus, modifications
should be made to $W_1$ if the breaking of $G$ to $H$ leads to the appearance of topological defects such as monopoles, strings
or domain walls. For instance, the breaking of $SU(4)_c\times SU(2)_L\times SU(2)_R$ \cite{patisalam} to the MSSM by fields belonging to
$\phi(\overline{4},1,2)$, $\overline{\phi}(4,1,2)$ produces magnetic monopoles that carry two quanta of Dirac magnetic charge
\cite{Lazarides:1980cc}. As shown in \cite{jeannerot}, one simple resolution of the monopole problem is achieved by supplementing $W_1$
with a non-renormalizable term consistent with $U(1)$ R-symmetry, 
whose presence enables an inf\mbox{}lationary trajectory along which the gauge symmetry is broken. Thus,
the magnetic monopoles are inf\mbox{}lated away. The part of the superpotential relevant for inf\mbox{}lation is given by

\begin{equation} \label{super2}
W_2=\kappa S(\overline{\phi}\phi-M^{2})-\beta\frac{S(\overline{\phi}\phi)^{2}}{M^{2}_{S}}\,,
\end{equation}

\noindent where $M$ is comparable to the GUT scale, $M_{S}\sim 5\times10^{17}$ GeV is a superheavy cutoff
scale, and the dimensionless coefficient $\beta$ is of order unity. 

Remarkably, the
inf\mbox{}lationary potential including one loop radiative corrections and the leading SUGRA correction is
obtained from Eq.\,(\ref{loop}) by substituting $\mathcal{N}=2$, $m^{2}=M^{2}(1/4\xi-1)$, $\sigma=\sqrt{2}|S|$, with
$\xi=\beta M^{2}/\kappa M^{2}_{S}$ and $z=x^{2}=\sigma^{2}/m^{2}$ \cite{jeannerot}:

\begin{eqnarray}
V_2 & \approx & \kappa^{2}m^{4}\bigg[1+\frac{\kappa^{2}}{16\pi^{2}} \Big(
2\ln\frac{2\kappa^{2}\sigma^{2}}{\Lambda^{2}}+(z+1)^{2}\ln(1+z^{-1})+ \nonumber \\
& &\hspace{4.5cm} +(z-1)^{2}\ln(1-z^{-1})\Big) +\frac{1}{8}\sigma^4\bigg]{}\,.
\end{eqnarray}

\noindent For $\kappa\lesssim0.01$ 
the SUGRA correction is negligible, and one obtains Eq.\,(\ref{quad}) with $\mathcal{N}$
replaced by 2, Eq.\,(\ref{m_kap}) with $\mathcal{N}$ replaced by 4, and $M$ replaced by $m$ in both equations. 
The dependence of the spectral index on $\kappa$ is depicted in Fig.\,5. The tensor to scalar ratio $r\lesssim10^{-4}$, 
while $\textrm{d}n_{s}/\textrm{d}\ln k\lesssim10^{-3}$. 

The vev
$v_{0}=\big|\langle\phi\rangle\big|=\big|\langle\overline{\phi}\rangle\big|$ at the SUSY minimum is given by \cite{jeannerot}

\begin{equation}
\left(\frac{v_{0}}{M}\right)^{2}=\frac{1}{2\xi}\left(1-\sqrt{1-4\xi}\right)\,,
\end{equation}

\noindent and is $\sim10^{16}-10^{17}$ depending on $\kappa$ (Fig.\,6). Requiring $v_0$ to be 
$\lesssim M_{GUT}\simeq2.86\times10^{16}$ GeV restricts $\kappa$ to $\lesssim0.01$ for $\beta=1$. However,
if one allows a smaller value for the effective cutoff scale which controls the non-renormalizable terms in the theory, 
say $M_S/\sqrt{\beta}\approx2\times10^{17}$ for $\beta=6$, then $v_0$ remains below $M_{GUT}$ 
even for $\kappa$ close to 0.1 (Fig.\,6). 

\section{Smooth GUT Inflation}
A variation on the inf\mbox{}lationary scenarios in sections 2 and 3 is obtained by imposing a $Z_{2}$ symmetry on the superpotential,
so that only even powers of the
combination $\phi\overline{\phi}$ are allowed \cite{smooth}. To leading order,

\begin{equation} \label{super3}
W_3=S\left(-\mu^2 +\frac{(\phi\overline{\phi})^{2}}{M^{2}_{S}}\right)\,,
\end{equation}

\noindent where the dimensionless parameters $\kappa$ and $\beta$ (see Eq.\,(\ref{super2})) are absorbed in $\mu$ and $M_S$
\cite{smooth}. 
This idea has also been implemented in a SUSY GUT model \cite{smooth2} based on the gauge group $SU(4)_c\times SU(2)_{L}\times SU(2)_{R}$ 
\cite{patisalam}. The inf\mbox{}lationary superpotential coincides with the `shifted' model of the previous section, 
except that the trilinear term is now absent.
The resulting scalar potential possesses two (symmetric) valleys of local minima which are suitable for inf\mbox{}lation and along which
the GUT symmetry is broken. The inclination of these valleys is already non-zero
at the classical level and the end of inf\mbox{}lation is smooth, in contrast to inf\mbox{}lation based on the superpotential $W_1$ 
(Eq.\,(\ref{super})). An important consequence is that, as in the case of shifted GUT inf\mbox{}lation, potential problems associated with primordial
topological effects are avoided.

The common vev at the SUSY minimum $M=\big|\langle\phi\rangle\big|=\big|\langle\overline{\phi}\rangle\big|=
(\mu\,M_S)^{1/2}$. For $\sigma^{2}\gg M^2$, the inf\mbox{}lationary potential is given by

\begin{equation} \label{v3}
V_3\approx\mu^{4}\left[1-\frac{2}{27}\frac{M^4}{\sigma^{4}}+\frac{1}{8}\sigma^4\right]\,.
\end{equation}

\noindent where the last term arises from the canonical SUGRA correction. If we set $M$ equal to the SUSY GUT scale $M_G$,
we get $\mu\simeq1.8\times10^{15}$ GeV and $M_S\simeq4.6\times10^{17}$ GeV. (Note that, if we express Eq.\,(\ref{super3}) in terms of
the coupling parameters $\kappa$ and $\beta$, these values correspond to $\kappa\sim O(\mu^2/M^2_{GUT})\sim 10^{-2}$ and $\beta\simeq1$.)
The value of the field $\sigma$ is
$1.3\times10^{17}$ GeV at the end of inf\mbox{}lation (corresponding to $\eta=-1$) and is $\sigma_Q=2.7\times10^{17}$ GeV at horizon exit. 
In the absence of the SUGRA correction (which is small for $M\lesssim10^{16}$ GeV), $\sigma\propto M^{2/3}\,M_P^{1/3}$, 
$(\delta T/T)_Q\propto M^{10/3}/(M^2_S\,M_P^{4/3})$ and the spectral index is given by \cite{smooth}

\begin{equation}
n_{s}\simeq1-\frac{5}{3N_{Q}}\simeq0.97\,,
\end{equation}

\noindent a value which coincides with the prediction of some D-brane inf\mbox{}lation models \cite{brane}. This may not be surprising
since, in the absence of SUGRA correction, the potential $V_3$ (Eq.\,(\ref{v3})) has a form familiar from D-brane inf\mbox{}lation.
The SUGRA correction raises $n_{s}$ from 0.97 to 1.0 
for $M\sim10^{16}$ GeV, and above unity for $M\gtrsim2\times10^{16}$ (Fig.\,7).

\section{Gravitino Constraint on $\kappa$}

 We have seen that the dimensionless superpotential coupling $\kappa$ in Eq.\,(\ref{super})
 satisfies the constraint $\kappa\lesssim0.1$, so that the scalar spectral index
 does not exceed unity by much on smaller scales ($k\gtrsim.05\,\textrm{Mpc}^{-1}$) from
 SUGRA corrections. A somewhat more stringent upper bound on $\kappa$ can
 be realized in an $SO(10)$ model from consideration of the inf\mbox{}laton decay
 into right-handed neutrinos, taking into account the gravitino constraint $T_r\lesssim
 10^{10}$ GeV on the reheat temperature. Consider the $SO(10)$ superpotential
 couplings

\begin{equation} \label{couplings}
\frac{1}{m_p}\gamma_{ij}\overline{\phi}\overline{\phi}16_i 16_j
\end{equation}
  
\noindent which provide large masses for the right-handed neutrinos after symmetry
 breaking (say of $SO(10)$ to $SU(5)$). Here $16_{i\textrm{,}\,j}$ ($i$, $j=1,2,3$) denote the
 three chiral families, and $\phi$, $\overline{\phi}$ vevs break the gauge symmetry.

 It was shown in \cite{lss} that the reheat temperature for the model discussed in section 2 can be approximated as

\begin{equation} \label{reheat}
T_r\sim\frac{1}{12}\,\sqrt{y_Q}\,M_i\,
\end{equation}

 \noindent where $y_Q$ is defined in Eq.\,(\ref{yq}), and $M_i$ denotes the heaviest right
 handed neutrino that satisfies $2M_i < m_{\mathrm{infl}}$, with

\begin{equation} \label{inflaton}
m_{\mathrm{infl}} = \sqrt{2} \kappa M\,.
\end{equation}

\noindent From Eq.\,(\ref{couplings}) we expect the heaviest neutrino to have a mass of around
$10^{14}$ GeV or so, which is in the right ball park to provide via the
seesaw a mass scale of about .05 eV to explain the atmospheric neutrino
anomaly through oscillations \cite{kyae}. From Eq.\,(\ref{reheat}), assuming $y_Q$ of order unity,
we conclude that $M_i$ should not be identified with the heaviest
right handed neutrino, otherwise $T_r$ would be too high. Thus, we require that \cite{Lazarides:1999rt}

\begin{equation}             
\frac{m_{\mathrm{infl}}}{2}\lesssim M_3=\frac{2M^2}{m_P}\quad(\textrm{with}\ \gamma_3=1)\,.
\end{equation}

\noindent Using Eq.\,(\ref{inflaton}) and Eq.\,(\ref{m_kap}), this implies $y_Q\lesssim\sqrt{\mathcal{N}N_Q}/\pi$, which corresponds to 
$\kappa\lesssim0.015$ (0.008) for $\mathcal{N}=16$ ($\mathcal{N}=2$).

\section{Conclusion}
Motivated by the prospects of much more precise data becoming available in the not too distant future, we have explored the
predictions of the scalar spectral index $n_s$, $\textrm{d}n_{s}/\textrm{d}\ln k$ and the tensor to scalar ratio $r$ for three
classes of closely related supersymmetric models. A characteristic feature shared by all these models is that inf\mbox{}lation becomes an 
integral part of realistic supersymmetric grand unified theories, with the grand unification scale playing an essential role.
If the grand unified theory is identified with $SO(10)$ or one of its rank five subgroups, then leptogenesis also becomes an
integral part of the setup. We find that the inclusion of SUGRA corrections can give rise to a blue spectrum
\cite{panagio,linde}, with the upper bound on $n_s$ (say $\leqslant1.0$ for $k=0.05$ Mpc$^{-1}$) 
providing an upper bound on $\kappa$ that is of order $10^{-1}\textrm{--}10^{-2}$ depending on the model.
The quantity $\textrm{d}n_{s}/\textrm{d}\ln k$ is always small, of order $10^{-3}$
or less \cite{kyaeet.,kawasaki}. 

\vspace{0.5cm}
\noindent {\bf Acknowledgments}
\\ \noindent
We thank G. Lazarides and S. Khalil for useful discussions. This work was supported by DOE under contract number DE-FG02-91ER40626.

\pagebreak

\begin{figure}[t]
\psfrag{N=16}{\scriptsize{$\mathcal{N}=16$}}
\psfrag{N=2}{\scriptsize{$\mathcal{N}=2$}}
\includegraphics[angle=0, width=13.5cm]{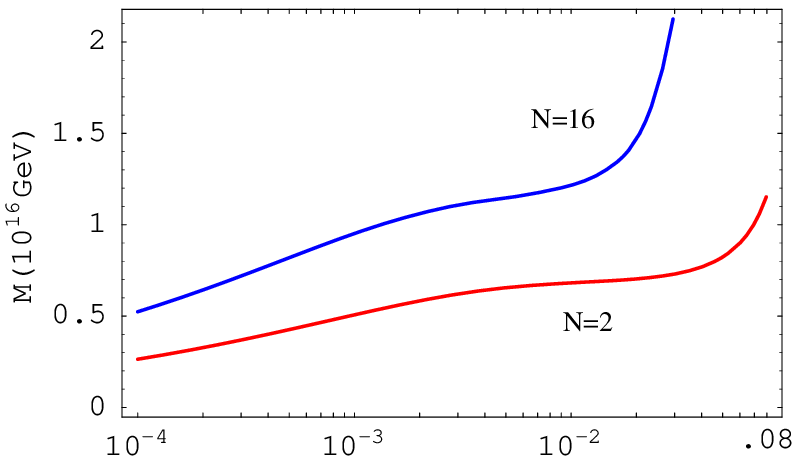}
\vspace{-.8cm}
\begin{center}
{\large \qquad $\kappa$}
\end{center}
\caption{\sf The gauge symmetry breaking scale $M$ as a function of the coupling constant $\kappa$.
$\mathcal{N}=16\ (2)$ corresponds to the breaking $SO(10)\to SU(5)$ and $SU(2)_L\times SU(2)_R\times U(1)_{B-L}\to
SU(2)\times U(1)$ respectively.}
\label{mvskappa}
\end{figure}

\begin{figure}[hbp]
\psfrag{N=16}{\scriptsize{$\mathcal{N}=16$}}
\psfrag{N=2}{\scriptsize{$\mathcal{N}=2$}}
\includegraphics[angle=0, width=13.5cm]{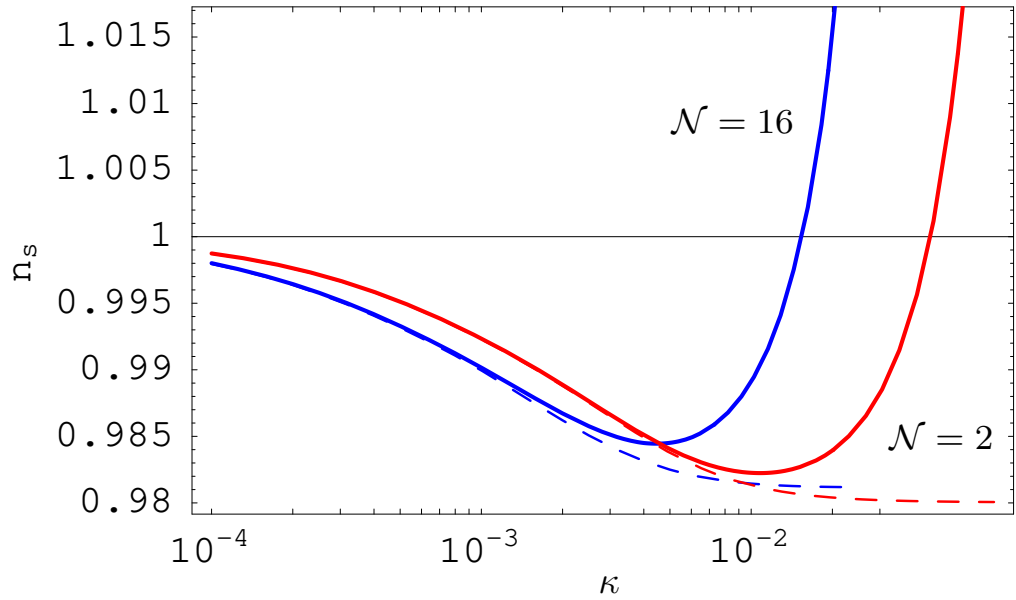}
\vspace{-1cm}
\begin{center}
{\large \qquad $\kappa$}
\end{center}
\caption{\sf The spectral index $n_s$ at $k=0.05 \ \textrm{Mpc}^{-1}$ as a function of the coupling constant $\kappa$
(dashed line--without SUGRA correction, solid line--with SUGRA correction).}
\label{nvskappa}
\end{figure}

\begin{figure}[htb]
\includegraphics[angle=0, width=14cm]{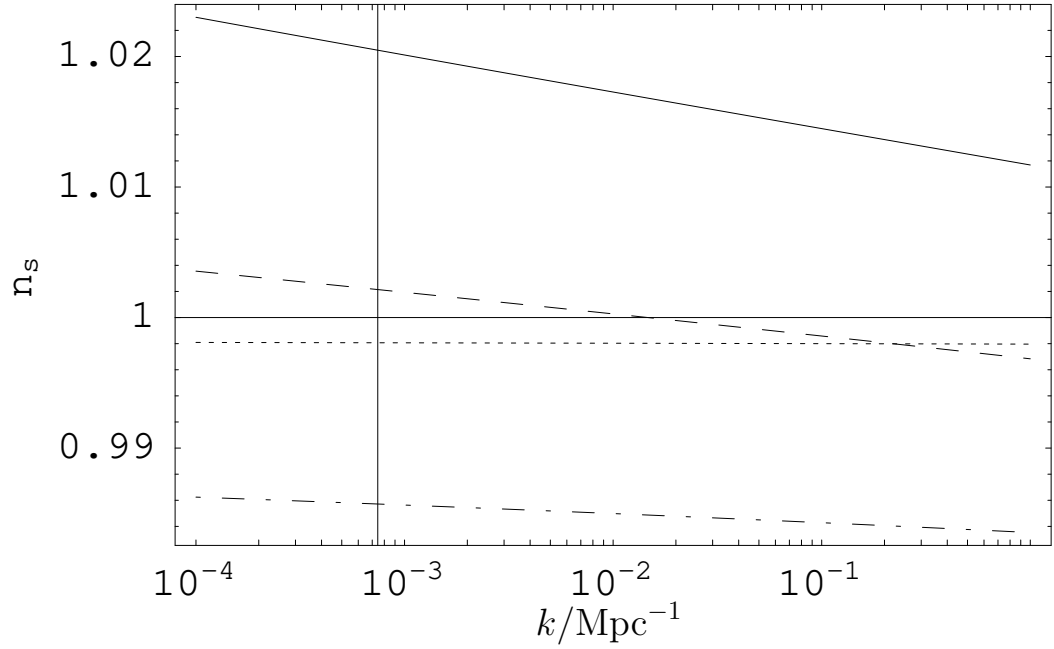}
\vspace{-.8cm}
\begin{center}
{\large \qquad $k/\textrm{Mpc}^{-1}$}\end{center}
\caption{\sf The spectral index $n_s$ as a function of the wavenumber $k$ $(\mathcal{N}=16)$: \newline
$\kappa=0.004$ (dot-dashed), $1\times10^{-4}$ (dotted), $0.015$ (dashed), $0.02$ (solid).}
\label{nvskappa2}
\end{figure}

\begin{figure}[htb]
\psfrag{N=16}{\scriptsize{$\mathcal{N}=16$}}
\psfrag{N=2}{\scriptsize{$\mathcal{N}=2$}}
\includegraphics[angle=0, width=14cm]{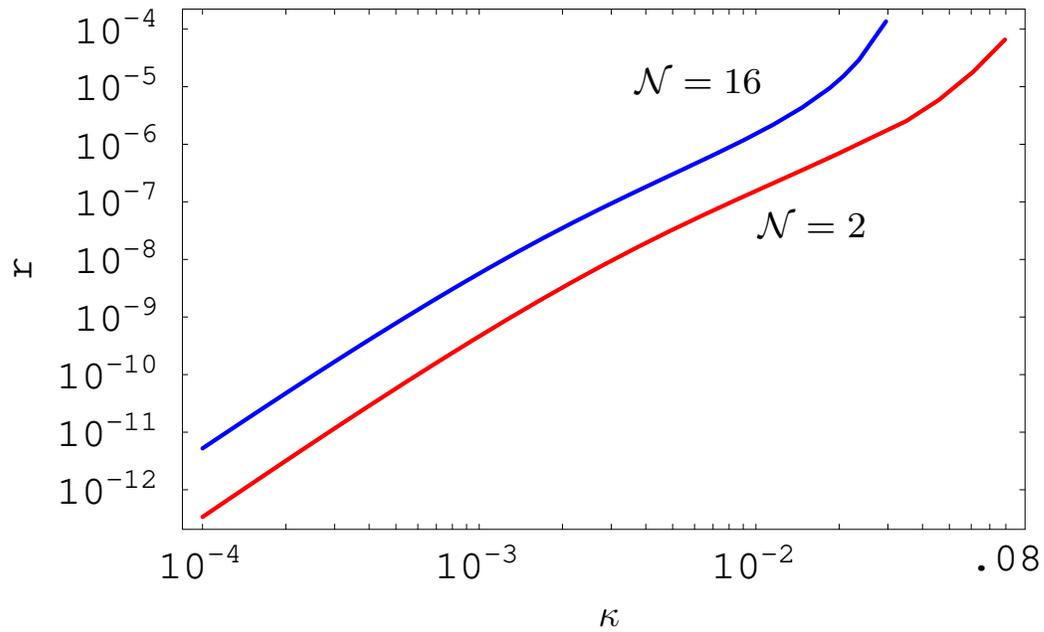}
\vspace{-.8cm}
\begin{center}
{\large \qquad $\kappa$}
\end{center}
\caption{\sf The tensor to scalar ratio $r$ as a function of the coupling constant $\kappa$.}
\label{rvskappa}
\end{figure}


\begin{figure}[htb]
\includegraphics[angle=0, width=13cm]{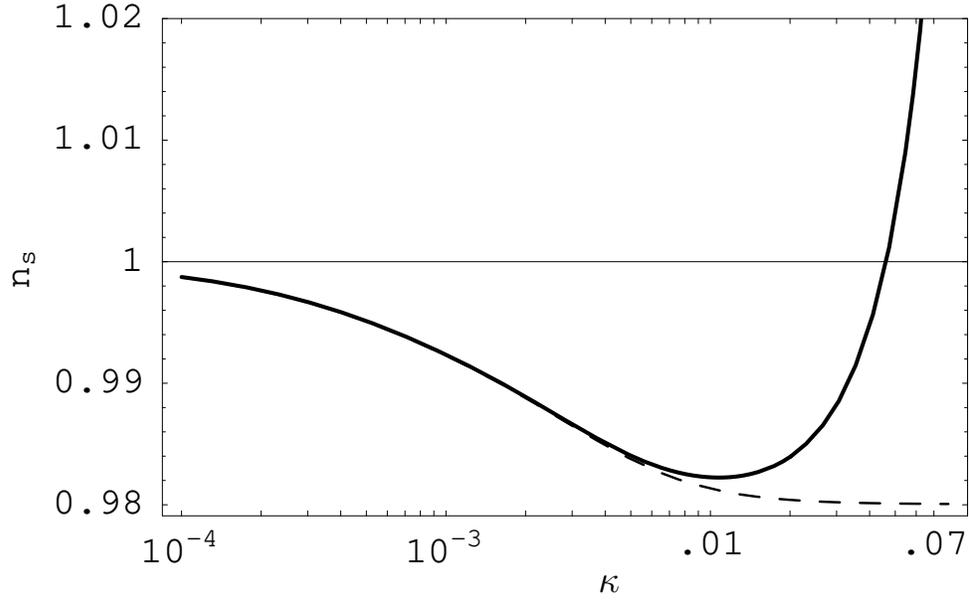}
\vspace{-.8cm}
\begin{center}
{\large \qquad $\kappa$}
\end{center}
\caption{\sf The spectral index $n_s$ at $k=0.05 \ \textrm{Mpc}^{-1}$ as a function of the coupling constant $\kappa$
for shifted GUT inf\mbox{}lation (dashed line--without SUGRA correction, solid line--with SUGRA correction).}
\label{nvskappa442}
\end{figure}

\begin{figure}[htb]
\psfrag{beta=1}{\scriptsize{$\beta=1$}}
\psfrag{beta=6}{\scriptsize{$\beta=6$}}
\includegraphics[angle=0, width=13cm]{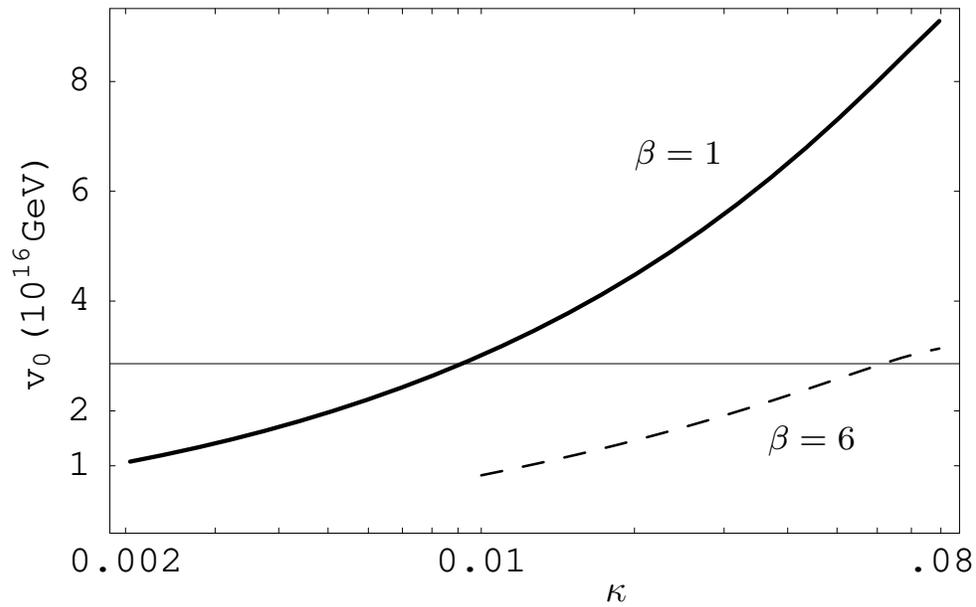}
\vspace{-.8cm}
\begin{center}
{\large \qquad $\kappa$}
\end{center}
\caption{\sf The gauge symmetry breaking scale $v_0$ in shifted GUT inflation as a function of the coupling constant $\kappa$ for $\beta=1$ (solid line) 
and $\beta=6$ (dashed line).}
\label{mvskappa442}
\end{figure}


\begin{figure}[t]
\includegraphics[angle=0, width=13cm]{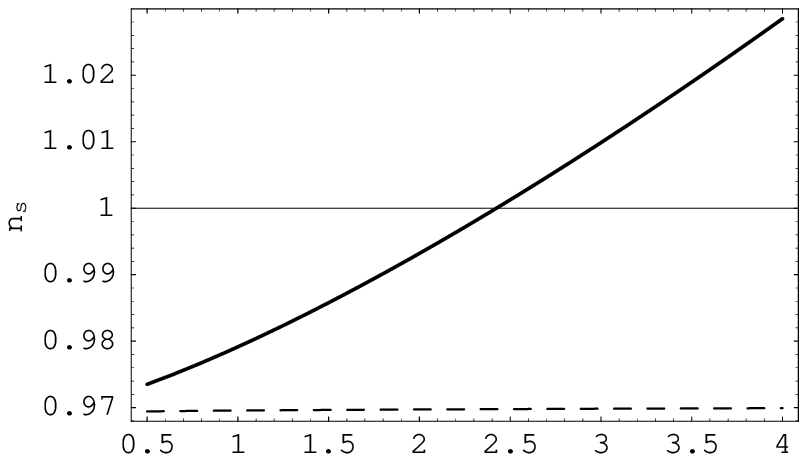}
\vspace{-.8cm}
\begin{center}
{\large \qquad $M\,(10^{16}\,\textrm{GeV})$}
\end{center}
\caption{\sf The spectral index $n_s$ at $k=0.05 \ \textrm{Mpc}^{-1}$ as a function of the gauge symmetry breaking
scale $M$
for smooth GUT inf\mbox{}lation (dashed line--without SUGRA correction, solid line--with SUGRA correction).}
\label{nvsksmooth}
\end{figure}


\begin{thebibliography}{99}
\begin{small}
\bibitem{dvaliet.}G. Dvali, Q. Shafi, and R. Schaefer, Phys. Rev. Lett. \textbf{73}, 1886 (1994) \\ \indent [hep-ph/9406319].

\bibitem{lazarides} For a comprehensive review, see G. Lazarides, Lect.Notes Phys. \textbf{592}, 351 (2002) 
\\ \indent [hep-ph/0111328].

\bibitem{wmap}D. N. Spergel \emph{et.al.}, astro-ph/0302209; H. V. Peiris \emph{et.al.}, astro-ph/0302225.

\bibitem{kyaeet.} B. Kyae, and Q. Shafi, astro-ph/0302504.

\bibitem{patisalam} J. C. Pati, and A. Salam, Phys. Rev. \textbf{D10}, 275 (1974).

\bibitem{king}
S.~F.~King and Q.~Shafi,
Phys.\ Lett.\ {\bf B422}, 135 (1998)
[hep-ph/9711288];
\\ \noindent I.~Antoniadis, and G.~K.~Leontaris,
Phys.\ Lett.\ {\bf B216}, 333 (1989);
\\ \noindent I.~Antoniadis, G.~K.~Leontaris, and J.~Rizos,
Phys.\ Lett.\ {\bf B245}, 161 (1990). 

\bibitem{jeannerot} R. Jeannerot, S. Khalil, G. Lazarides, and Q. Shafi, JHEP \textbf{10}, 12
(2000) \\ \indent [hep-ph/0002151].

\bibitem{lss}G. Lazarides, R. K. Schaefer and Q. Shafi, Phys. Rev. \textbf{D56}, 1324 (1997) \\
\indent [hep-ph/9608256].

\bibitem{dls}G. Dvali, G. Lazarides, and Q. Shafi, Phys. Lett. \textbf{B424}, 259 (1998) \\
\indent [hep-ph/9710314].

\bibitem{panagio}C. Panagiotakopoulos, Phys. Rev. \textbf{D55}, 7335 (1997) [hep-ph/9702433];
\\ \indent W.~Buchmuller, L.~Covi, and D.~Delepine,
Phys.\ Lett.\ {\bf B491}, 183 (2000) \\ \indent [hep-ph/0006168]. 

\bibitem{linde} A. Linde, and A. Riotto, Phys. Rev. \textbf{D56}, 1841 (1997) [hep-ph/9703209].

\bibitem{kawasaki}
M.~Kawasaki, M.~Yamaguchi, and J.~Yokoyama,
hep-ph/0304161.


\bibitem{covi}
L.~Covi, G.~Mangano, A.~Masiero, and G.~Miele,
Phys.\ Lett.\ {\bf B424}, 253 (1998)
\\ \indent [hep-ph/9707405].  

\bibitem{smooth} G. Lazarides, and C. Panagiotakopoulos, Phys.\ Rev.\ {\bf D52}, 559 (1995) \\ \indent [hep-ph/9506325];
\\ \indent G. Lazarides, C. Panagiotakopoulos, and N. D. Vlachos, Phys.\ Rev.\ {\bf D54}, 1369 (1996) [hep-ph/9606297].

\bibitem{copeland} E. J. Copeland, A. R. Liddle, D. H. Lyth, E. D. Stewart, and D. Wands, 
\\ \indent Phys. Rev. \textbf{D49}, 6410 (1994).


\bibitem{cobe} G. F. Smoot \emph{et.al.}, Astrophys. J. Lett. \textbf{396}, L1(1996); \\ \indent C. L. Bennett
\emph{et.al.}, Astrophys. J. Lett. \textbf{464}, 1 (1996); \\ \indent 
E. F. Bunn, A. R. Liddle, and M. White, Phys. Rev. \textbf{D54}, 5917
(1996) \\ \indent [astro-ph/9607038].

\bibitem{kyae} B. Kyae and Q. Shafi, hep-ph/0212331.

\bibitem{ll}A. R. Liddle, and D. H. Lyth, Phys. Lett. \textbf{B291}, 391 (1992) [astro-ph/9208007];\\ \indent
Phys. Rep. \textbf{231}, 1 (1993) [astro-ph/9303019]

\bibitem{gravitino} M. Yu. Khlopov, and A. D. Linde, Phys. Lett. {\bf B138},
265 (1984);\\J. Ellis, J. E. Kim, and D. Nanopoulos,
Phys. Lett. {\bf B145}, 181 (1984).

\bibitem{ls} G. Lazarides, and Q. Shafi, Phys. Lett. {\bf B258}, 305 (1991).

\bibitem{lsv} G. Lazarides, Q. Shafi, and N. D. Vlachos, Phys. Lett {\bf B427}, 53 (1998) \\
\indent [hep-ph/9706385]; \\ \indent G.~Lazarides, and Q.~Shafi,
Phys.\ Rev.\ {\bf D58}, 071702 (1998)
[hep-ph/9803397]. 

\bibitem{Pati:2002pe}
J.~C.~Pati,
hep-ph/0209160; \\ \indent T. Asaka, hep-ph/0304124.

\bibitem{Panagiotakopoulos:1997ej}
C.~Panagiotakopoulos,
Phys.\ Lett.\  {\bf B402}, 257 (1997)
[hep-ph/9703443].

\bibitem{Lazarides:1998zf}
G.~Lazarides and N.~Tetradis,
Phys.\ Rev.\  {\bf D58}, 123502 (1998)
[hep-ph/9802242].

\bibitem{Hamaguchi:2002vc}
K.~Hamaguchi,
hep-ph/0212305.

\bibitem{lepto} M. Fukugita, and T. Yanagida,
Phys. Lett. {\bf B174}, 45 (1986). \\ \indent For a recent discussion of thermal leptogenesis and additional references see
W. Buchmuller, P. diBari, and M. Plumacher, hep-ph/0302092. 
Note that in the models we are discussing thermal leptogenesis is possible
if we allow the reheat temperature to be close to $10^{10}$ GeV.

\bibitem{seljak} U. Seljak, P. McDonald, and A. Makarov, astro-ph/0302571.

\bibitem{Lazarides:1980cc}
G.~Lazarides, M.~Magg, and Q.~Shafi,
Phys.\ Lett.\ {\bf B97}, 87 (1980).


\bibitem{smooth2} R. Jeannerot, S. Khalil, and G. Lazarides, Phys.\ Lett.\ {\bf B506}, 344 (2001)\\ \indent [hep-ph/0103229].

\bibitem{brane} G. Dvali, Q. Shafi, and S. Solganik, hep-th/0105203; \\
C. P. Burgess \emph{et. al.}, JHEP \textbf{0107}, 047 (2001) [hep-th/0105204]; \\
B. Kyae, and Q. Shafi, Phys. Lett. \textbf{B526}, 379 (2002) [hep-ph/0111101]; \\
C.~Herdeiro, S.~Hirano, and R.~Kallosh,
JHEP {\bf 0112}, 027 (2001)
\\ \indent [hep-th/0110271];\\
J.~H.~Brodie, and D.~A.~Easson,
hep-th/0301138.

\bibitem{Lazarides:1999rt}
G.~Lazarides, and N.~D.~Vlachos,
Phys.\ Lett.\  {\bf B459}, 482 (1999)
[hep-ph/9903511].

\end{small}
\end{thebibliography}
\end{document}